**Simple stochastic birth and death models of genome evolution:**

**Was there enough time for us to evolve?**

**Georgy P. Karev, Yuri I. Wolf, and Eugene V. Koonin**

National Center for Biotechnology Information, National Library of Medicine, National Institutes of Health, Bethesda, MD 20894

**Abstract**

We show that simple stochastic models of genome evolution lead to power law asymptotics of protein domain family size distribution. These models, called Birth, Death and Innovation Models (BDIM), represent a special class of balanced birth-and-death processes, in which domain duplication and deletion rates are asymptotically equal up to the second order. The simplest, linear BDIM shows an excellent fit to the observed distributions of domain family size in diverse prokaryotic and eukaryotic genomes. However, the stochastic version of the linear BDIM explored here predicts that the actual size of large paralogous families is reached on an unrealistically long timescale. We show that introduction of non-linearity, which might be interpreted as interaction of a particular order between individual family members, allows the model to achieve genome evolution rates that are much better compatible with the current estimates of the rates of individual duplication/loss events.

# 1. Introduction.

Power distributions appear in an astonishing variety of fundamentally different contexts. These characteristic curves, that have been originally introduced as the Pareto law in economics [Pareto, 1897] and the Zipf law in linguistics [Zipf, 1949] describe the distribution of the number of links between documents in the Internet, the population of towns, the number of species that become extinct within a year, the number of sexual and other contacts between people, and numerous other quantities [Barabasi, 2002; Mendes, 2003]. Mathematically, these distributions are based on the negative power law: $P(i) \cong ci^{-\gamma}$ where $P(i)$ is the frequency of nodes with exactly $i$ connections or sets with exactly $i$ members, $\gamma$ is a parameter which typically assumes values between 1 and 3, and $c$ is a normalization constant. Obviously, in double-logarithmic coordinates, the plot of $P$ as a function of $i$ is close to a straight line with a negative slope.

The advent of genome sequencing brought about a surge in power law analyses in the field of genomics [Searls, 2002 ; Koonin, 2002; Luscombe, 2002]. The "dominance by a selected few" [Luscombe, 2002] embodied in the power laws has been noticed in the distribution of the number of transcripts per gene, the number of interactions per protein, the number of genes or pseudogenes in paralogous families, the number of connections per node in metabolic networks, and other quantities that can be obtained by genome analysis [Qian, 2001; Jeong, 2001; Jeong, 2000; Luscombe, 2002].

A recent detailed study showed that the distributions of several genome-related quantities claimed to follow power laws, e.g., the number of transcripts per gene, are better described by the so-called generalized Pareto function: $P(i) = c(i+a)^{-\gamma}$ where $a$ is an additional parameter [Kuznetsov, 2002 ; Kuznetsov, 2001]. Obviously at large $i$ ($i \gg a$), a generalized Pareto distribution is indistinguishable from a power law, but at small $i$, it deviates significantly, with the magnitude of the deviation depending on $a$. Notably, unlike power law distributions, generalized Pareto distributions do not show scale-free properties over the entire range of the argument.

The question that emerges when the same mathematical structure appears in apparently unrelated contexts is: are these formal similarities coincidental and superficial or do they reflect a deep connection at the level of evolutionary mechanisms [Sole, 1996; Gisiger, 2001] ? The applicability of the preferential attachment principle to the evolution of systems with power law type distributions suggests that the latter view is closer to the truth. More specifically, however, the epistemological value of the analysis of these distributions seems to lie in the connection between specific forms of the distributions with distinguishable evolutionary models. Such evolutionary modeling has been applied to genome-specific distributions of paralogous family size, [Huynen, 1998; Rzhetsky, 2001; Qian, 2001; Karev, 2002] the distribution of folds and families in the entire protein universe [Dokholyan, 2002], and protein-protein interaction networks [Pastor-Satorras, 2003; Wagner, 2003].

In our previous work [Karev, 2002; Koonin, 2002], we undertook such a mathematical analysis using a simple model of evolution with duplication (birth), elimination (death) and *de novo* emergence (innovation) of a domain as elementary processes (hereinafter BDIM, after birth-death-innovation model). We proved that the power asymptotic appears if, and only if, birth and death rates of domains in families of sufficiently large size are balanced (asymptotically equal up to the second order) and that any power asymptotic with $\gamma \neq 1$ can appear only if the per domain duplication/deletion rates depend on the size of a domain family. We applied the developed formalism to the analysis of the size distributions of domains in individual prokaryotic and eukaryotic genomes and showed a good fit between these data and a particular form of the model, the second-order balanced linear BDIM.

Here, we examine the non-deterministic version of BDIM and concentrate on the stochastic characteristics of the system, such as the probability of the formation a family of the given size before extinction and the mean times of formation and extinction of a family of a given size. We first investigate these issues within the framework of the linear $2^{nd}$ order balanced birth-and-death process. Given the published estimates of the rates of gene duplication and loss, we conclude that this version of BDIM, which fits well the stationary distributions of family sizes for different genomes, predicts completely unrealistic times for reaching the observed sizes of the largest domain families. We suggest a non-linear modification of the initial model whose stochastic characteristics are more realistic.

**2. Definitions, assumptions and empirical data**

We treat a genome as a "bag" of genes (gene fragments), coding for protein domains, which we will simply call *domains* for brevity (see [Karev, 2002 ] for additional details and rationale). Domains are treated as independently evolving units disregarding the dependence between domains that tend to belong to the same multidomain protein. Each domain is considered to be a member of a family, which may have one or more members. Three classes of elementary events are considered:

  i) domain *birth* which generates a new member in the same family as a result of gene is duplication
  ii) domain *death*, i.e., inactivation and/or deletion, and
  iii) *innovation* which generates a new family with one member. Innovation may occur via domain evolution from a non-coding sequence or a sequence of a non-globular protein, via horizontal gene transfer from another species, or via radical modification of a domain following a duplication. The rates of elementary events are considered to be independent of time (only homogeneous models are considered) and of the nature (structure, biological function, and other features) of individual families.

The data on the size of domain families in sequenced genomes were from the previous work [Karev, 2002 ]. Briefly, the domains were identified by comparing the CDD library of position-specific scoring matrices (PSSMs), which includes the domains from the Pfam and SMART databases, to the protein sequences from completely sequenced eukaryotic and prokaryotic genomes (http://www.ncbi.nlm.nih.gov/entrez/query.fcgi?db=Genome) using the RPS-BLAST program [Marchler-Bauer, 2002 ].

In a finite genome, the maximal number of domains in a family obviously cannot exceed the total number of domains and, in reality, is probably much smaller. Let $N$ be the maximal possible number of domain family members (note that almost all of the results below are valid with $N=\infty$ under certain well defined conditions, which provide the existence of the ergodic distribution of the birth-and-death process). We also consider "virtual" families consisting of 0 domains. In the model, newborn domains are extracted from this class and dead domains return to it. In the previous work [Karev, 2002], we examined exclusively the deterministic version of BDIM. Introduction of the 0 class "closes" the model and, accordingly, transforms it into a Markov process. This provides for the possibility to explore the stochastic properties of the system. In these stochastic models, innovation was not introduced explicitly but is implied in the form of emergence of domains from the 0 class.

We assume that: i) time is continuous and more than one elementary event is unlikely to occur in a short time interval (probability that more than one event occurs during an interval $\Delta t$ is $o(\Delta t)^2$), ii) all elementary events are independent of each other, and iii) the rates of domain birth and death depend on family size only. Let $p_i(t)$ be the frequency of a domain family of size $i$. Then $p_i(t)$ satisfy a well known system of forward Kolmogorov equations for birth-and-death process (see, e.g., [Anderson, 1991; Grimmett, 1992]:

$$d\, p_0(t)/dt = -\lambda_0\, p_0(t) + \delta_1 p_1(t),$$

$$d\, p_1(t)/dt = \lambda_0\, p_0(t) - (\lambda_1 + \delta_1) p_1(t) + \delta_2 p_2(t),$$

$$d\, p_i(t)/dt = \lambda_{i-1} p_{i-1}(t) - (\lambda_i + \delta_i) p_i(t) + \delta_{i+1} p_{i+1}(t) \text{ for } 1<i<N, \tag{2.1}$$

$$d\, p_N(t)/dt = \lambda_{N-1} p_{N-1}(t) - \delta_N\, p_N(t).$$

It is known that model (2.1) has a unique stationary ergodic distribution $p_0, ..., p_N$ defined by the equalities $dp_i(t)/dt=0$ for $0 \leq i \leq N$, such that

$$p_i = p_0 \prod_{j=1}^{i} (\lambda_{j-1}/\delta_j) \text{ for all } i=1,\ldots N, \tag{2.2}$$

$$p_0 = [1 + \sum_{i=1}^{N} \prod_{j=1}^{i} (\lambda_{j-1}/\delta_j)]^{-1}.$$

We will consider also the variant of the model (2.1) without the 0-state:

$dp_1(t)/dt = -\lambda_1 p_1(t) + \delta_2 p_2(t),$

$d p_i(t)/dt = \lambda_{i-1} p_{i-1}(t) - (\lambda_i + \delta_i) p_i(t) + \delta_{i+1} p_{i+1}(t)$ for $1 < i < N$, (2.3)

$d p_N(t)/dt = \lambda_{N-1} p_{N-1}(t) - \delta_N p_N(t).$

This model describes the evolution of the size of a domain family that includes an indispensable (essential) gene and is not allowed to go extinct. For this model, the ergodic distribution is

$$p_i = p_1 \prod_{j=2}^{i} (\lambda_{j-1}/\delta_j) \text{ for all } i=1,\ldots N,$$ (2.4)

$$p_1 = [1 + \sum_{i=2}^{N} \prod_{j=2}^{i} (\lambda_{j-1}/\delta_j)]^{-1}.$$

Mathematically, systems (2.1) and (2.3) describe the state probabilities of well-known birth-and-death processes with a finite number of states and reflecting boundaries. Although this classical process has been studied in detail, it has not been previously noticed that it is a natural source of the power-law distributions.

**3. Ergodic distribution of the model and the power asymptotics**

The ergodic distribution (2.2) or (2.4) is globally stable and is approached exponentially with respect to time from any initial state.

Let us define a function $\chi(i) = \lambda_{i-1}/\delta_i$, which describes the relation between the flow into and the flow from the state $i$. We consider only functions that grow no faster than a power function, i.e., $\chi(i) \leq i^s$ at $i \to \infty$ for a real $s$ (this is not a serious restriction because the most realistic situations correspond to the case of $s=0$; see Theorem 1 below).

The ergodic distribution is completely defined by the asymptotic behavior of the function χ(*i*). Let us suppose that, for large *i*, the following expansion is valid:

$$\chi(i) = \lambda_{i-1}/\delta_i = i^s \theta (1 - \gamma/i + O(1/i^2)) \quad (3.1)$$

where *s* and γ are real numbers and θ is positive. The following main assertions regarding different orders of balance in BDIM (as defined in the formulation of the theorem below) were proved previously [Karev, 2002].

**Theorem 1.** (i) *if s≠0 (non-balanced BDIM), then* $p_i \sim \Gamma(i)^s \theta^i i^{-\gamma}$ *where* Γ(*i*) *is the Γ-function;*

(ii) *if s=0 and* θ≠1 *(first-order balanced BDIM), then* $p_i \sim \theta^i i^{-\gamma}$;

(iii) *if s=0;* θ=1 *and a≠0 (second-order balanced BDIM), then* $p_i \sim i^{-\gamma}$;

(iv) *if s=0;* θ=1 *and a=0 (high-order balanced BDIM), then* $p_i \sim 1$.

Mathematically, the relation (3.1) at *s*=0 is simply the Taylor expansion of χ(*i*) over 1/*i*; hence, theorem 1 asserts that, for a balanced BDIM, the asymptotic of the ergodic distribution is completely determined by the first two coefficients of this expansion and, for the second order balanced BDIM, the power γ of the asymptotical power distribution is exactly equal to the second coefficient of this expansion.

**Corollary 1**. *For a non-balanced BDIM with s=-1 and* θ<1, *and* γ=0, *the equilibrium frequencies* $p_i$ *follow the (truncated) Poisson distribution with parameter* θ, $p_i \sim \theta^i/i!$

**Corollary 2**. *For a first-order balanced BDIM with* θ<1,

a) *if* γ<1, *the equilibrium frequencies* $p_i$ *follow Pascal distribution with parameters* (1-γ, θ);

b) *if γ= 1, the equilibrium frequencies follow (truncated) logarithmic distribution with parameter* θ;

c) *if γ=0, the equilibrium frequencies follow geometric distribution with parameter* θ.

The following implication of Theorem 1 is of principal interest.

**Corollary 3**. *Equilibrium frequencies of a BDIM have a power asymptotic behavior if and only if the BDIM is second-order balanced.*

**Corollary 4**. *For high-order balanced BDIM, if $\lambda_{i-1}/\delta_i =1$ for all i, then the only possible distribution of equilibrium frequencies is uniform, $p_i = $ const for all i. Moreover, even if $\lambda_{i-1}/\delta_i=1+O(1/i^2)$, the equilibrium frequencies asymptotically have the uniform distribution.*

The non-balanced BDIM (i) at *s*<-1 or *s*>0 and high-order balanced BDIM (iv) are of little practical interest because the former results in an extremely sharply dropping (or rising) distribution, whereas the latter leads to a uniform distribution of domain family sizes. Neither of these cases is observed in real-world situations, so in what follows we consider only the balanced BDIM. Precise formulas for $p_i$ can be obtained for specific forms of $\lambda_i$ and $\delta_i$ (see (Karev, Wolf, Rzhetsky, Berezovskaya and Koonin, 2002a) for details) and several of these will be considered below.

In the simplest case, the per-domain birth and death rates do not depend on family size; thus, per-family birth and death rates are proportional to the number of domains in a family: $\lambda_i=\lambda i$ and $\delta_i=\delta i$, $\lambda$ and $\delta$ are positive constants (simple BDIM). A simple BDIM can be either first-order balanced ($\lambda \neq \delta$) or second-order balanced ($\lambda=\delta$). According to the theorem above, the equilibrium distribution of domain family sizes could be either $p_i \sim (\lambda/\delta)^i/i$ (truncated logarithmic distribution when $\lambda<\delta$) for the first-order balanced simple BDIM or $p_i \sim 1/i$ for the second-order balanced simple BDIM. If a power asymptotics with $\gamma \neq 1$ is observed, a simple BDIM has to be rejected as the underlying model of evolution.

If the per-family birth and death rates linearly depend on the number of domains in a family

$\lambda_i = \lambda(i+a)$, $\delta_i = \delta(i+b)$ for $i>0$, $a$ and $b$ are constants   (3.2)

(linear BDIM), the equilibrium distribution of domain family sizes is defined by the following formula:

$$p_i = p_0 \frac{\Gamma(1+b)}{\Gamma(1+a)} \frac{\lambda_0}{\lambda} \theta^i \frac{\Gamma(i+a)}{\Gamma(i+1+b)} \sim \theta^i i^{a-b-1}, \text{ where } \theta = \lambda/\delta. \quad (3.3)$$

A linear BDIM is, by definition, at least first-order balanced; if $\lambda = \delta$ ($\theta = 1$), the resulting second-order balanced linear BDIM has a power asymptotics with $\gamma = 1+b-a$. The total number of families and domains at equilibrium and the ratio of total birth rate to innovation rate formally depend on the maximum family size $N$. However, it can be shown that, for the real-world domain family size distributions, this dependence is very weak for large $N$.

A wide class of $\lambda_i$ and $\delta_i$ functions can be described or approximated in terms of a rational BDIM:

$$\lambda_i = \lambda P(i) = \lambda \prod_{k=1}^{n}(i+a_k)^{\alpha_k}, \quad \delta_i = \delta Q(i) = \delta \prod_{k=1}^{m}(i+b_k)^{\beta_k}. \quad (3.4)$$

In this case, the equilibrium frequencies have the following asymptotics

$$p_i \sim \Gamma(i)^\eta \theta^i i^\rho \quad (3.5)$$

where $\theta = \lambda/\delta$; $\eta = \sum_{k=1}^{n} \alpha_k - \sum_{k=1}^{m} \beta_k$ and $\rho = \sum_{k=1}^{n} a_k \alpha_k - \sum_{k=1}^{m} b_k \beta_k - \sum_{k=1}^{m} \beta_k$.

## 4. Linear BDIM and its applications

Previously, we applied BDIMs to approximate the observed distribution of protein domains in several prokaryotic and eukaryotic genomes by minimizing the $\chi^2$ value for the observed and predicted distributions. The simplest model that resulted in a good fit to the observed domain family size distributions for all analyzed genomes was the second-order balanced linear BDIM. For all analyzed genomes, $P(\chi^2)$ for this model was >0.05, i.e., no significant difference between the model predictions and the observed data was detected. The asymptotics of the distribution implied by the second-order balanced linear BDIM is a power law, with $\gamma = 1- a+b$ (*a* and *b* are the parameters of a linear BDIM). We observed that, for all analyzed genomes, $\gamma >1$ ($a<b$), which corresponds to "synergy" between domains in a family. In other words, small families appear to be less stable during evolution than large families, whereas members of large families have a greater likelihood of survival over long intervals of evolution. The linear BDIM adequately accommodates even the largest of the identified domain families. Lineage-specific expansion of paralogous families is well recognized as one of the principal modes of adaptation. Thus, it appears that, quantitatively, adaptive family expansion is within the framework of the BDIMs, although these models do not explicitly incorporate the notion of selection.

In what follows, we study the behavior of the stochastic linear BDIM in more details.

## 5. Probabilities of formation of families of different sizes for the linear BDIM

In is known [Dynkin, 1969 ; Bhattacharya, 1990] that the probability for the birth-and-death process to reach state *n* before reaching state 0 from an initial state $i>0$ is

$$p(i,n) = (1+\sum_{j=1}^{i-1} \prod_{k=1}^{j} \delta_k/\lambda_k)/(1+\sum_{j=1}^{n-1} \prod_{k=1}^{j} \delta_k/\lambda_k) \qquad (5.1)$$

In terms of the BDIM (2.1), this means that the probability of formation of a family of size *n* starting from a family of size *i* before getting to extinction is given by (5.1); in particular, the probability that a singleton expands to a family of size *n* before dying is

$$P_n = 1/(1+\sum_{j=1}^{n-1} \prod_{k=1}^{j} \delta_k/\lambda_k). \qquad (5.2)$$

For the linear 2nd order balanced BDIM, the probability that a singleton expands to a family of size $n$ before dying is

$$P_n = 1/(1+\frac{\Gamma(1+a)}{\gamma\Gamma(1+b)}(\frac{\Gamma(b+n+1)}{\Gamma(a+n)} - \frac{\Gamma(2+b)}{\Gamma(1+a)})) \qquad (5.3)$$

where $\gamma=1+b-a$.

Note that these probabilities have the power asymptotics for large $n$:

$$P_n \cong \frac{\gamma\Gamma(1+b)}{\Gamma(1+a)} n^{-\gamma}. \qquad (5.4)$$

with the same degree as the equilibrium frequencies of the families. The values of probabilities $P_n$ for different species are shown in Figure 1 and Table 2.

## 6. Mean time of extinction of a family of a given size for the linear BDIM

The random birth-and-death process (2.1) certainly visits the state 0 in the course of time; this means that any domain family will eventually get extinct (and then formally can be "reborn", returning from the 0-class). Let us compute the mean time of extinction of a family of the given size; the mean time of extinction of the largest family is the value of greatest interest.

Let us denote $W(n)=\inf\{t: X(t)=0| X(0)=n\}$ the time of the first passage of state 0 from the initial state $n$; $W(n)$ is a random variable for each $n$. The mean time of extinction of the family of initial size $n$, $E(W(n))$, can be calculated using the following formula (see, e.g., [Anderson, 1991; Bhattacharya, 1990]):

$$E(W(n))= \sum_{k=1}^{n} \sum_{i=k}^{N} (\lambda_k...\lambda_{i-1})/(\delta_k...\delta_i). \qquad (6.1)$$

For the linear 2nd order balanced BDIM, $E(W(n)) = 1/\lambda\, E_n$, where

$$E_n = \sum_{k=1}^{n} \frac{\Gamma(k+b)}{\Gamma(k+a)} \sum_{i=k}^{N} \frac{\Gamma(i+a)}{\Gamma(i+1+b)}. \qquad (6.2)$$

The plot of $E_n$ versus $n$ for different species is shown in Figure 2.

**7. The mean time of formation of a family of a given size for the linear BDIM**

Let us denote $T(i,n)=\inf\{t: X(t)=n|\ X(0)=i\}$ the time of the first passage of state $n$ from the initial state $i$; $T(i,n)$ is a random variable for each $i, n$. The mean time of the first passage for BDIM (2.1), $m(i;n)=E(T(i,n))$, can be calculated using the following formula (Bhattacharya, Waymire, ch.3):

$$m(i;n)= \sum_{m=i}^{n-1} \sum_{k=0}^{m} \delta_{k+1}\ldots\delta_m/(\lambda_k\ldots\lambda_m). \qquad (7.1)$$

It is convenient to explicitly write $m(1;n)$, the mean time required to reach state $n$ from state 1 in the form:

$$m(1,n)= m_0(n) + m_1(n),$$

where $m_0(n)= 1/\lambda_0 \sum_{m=1}^{n-1} \delta_1\ldots\delta_m/(\lambda_1\ldots\lambda_m),$

$$m_1(n)= \sum_{m=1}^{n-1} \sum_{k=1}^{m} \delta_{k+1}\ldots\delta_m/(\lambda_k\ldots\lambda_m). \qquad (7.2)$$

The term $m_0(n)$ is the mean time elapsed before the system leaves the 0 state for the last time and the term $m_1(n)$ is the mean time of formation of a family of size $n$ from a singleton after its last resurrection.

The values of $\lambda m_1(n)$ and $\lambda_0 m_0(n)$ for the second order balanced linear stochastic BDIM and for each of the genomes are rather close to each other. For example, $m(1,1151)$ $=1/\lambda\ 300665.09 + 1/\lambda_0\ 382994.665$ for *H. sapiens* (1151 is the size of the largest family encoded in the human genome according to our data). Hence, the relative input of the two terms to the total mean formation time depends on the values of $\lambda$ and $\lambda_0$, respectively. It can be easily shown that $M(n)$, the mean formation time from an essential singleton (see model (2.3)), is exactly equal to $m_1(n)$. In what follows we study only the mean time $M(n)$; for the linear BDIM

$$M(n) = 1/\lambda \, M_n = 1/\lambda \sum_{m=1}^{n-1} \left( \frac{\Gamma(b+m+1)}{\Gamma(a+m+1)} \sum_{k=1}^{m} \frac{\Gamma(a+k)}{\Gamma(b+k+1)} \right). \tag{7.3}$$

Next, let us consider the process of formation of families in more detail and find the mean time of formation, from an essential singleton, of a family of size $n$, $M(n)$. Plots of $M_n$ (the mean time of formation for $\lambda=1$) are shown on Figures 3a and 3b.

Let us note that $\lambda$ is an interior parameter of the model and cannot be equated with the actual average duplication rate, $r_{du}$, which can be estimated from empirical data. To connect these two values, one should take into account that $r_{du}$ is estimated as the average duplication rate per domain over the entire genome. As $\lambda_i/i$ is the duplication rate per domain in a family of size $i$, then, for model (2.3)

$$r_{du} = \sum_{i=1}^{N-1} p_i \lambda_i / i \tag{7.4}$$

where $N$ is the maximal family size in the given genome (note that the duplication rate in class $N$ is 0 by definition).

Then, after simple algebra, it can be shown that, for the linear 2$^{nd}$ order balanced model, the following coefficient, $c_{du}$, connects the model parameter $\lambda$ with the empirical estimate of the duplication rate:

$$c_{du} = r_{du}/\lambda = \sum_{i=1}^{N-1} \frac{\Gamma(a+i+1)}{i\Gamma(b+i+1)} \Big/ \sum_{j=1}^{N} \frac{\Gamma(a+j)}{\Gamma(b+j+1)}. \tag{7.5}$$

The values of $M_N$ and $c_{du}$ for different species are given in Table 2. The model parameter $\lambda$ actually is of the same order of magnitude as the mean duplication rate per domain for the linear BDIM.

Summarizing the results obtained for the stochastic characteristics of the linear BDIM, we found that, firstly, there is an extremely large difference between the times of formation and extinction of the largest families for some genomes, the latter being much more rapid. Secondly, and most importantly, the above connection between $\lambda$ and $r_{du}$ allows one to use the available conservative empirical estimates of duplications rates to express the mean family formation times in real time units instead of the dimensionless $1/\lambda$ units. These estimates, which were produced by counting the number of recent duplicates in three eukaryotic genomes and dividing this number by the estimated rate of silent nucleotide substitutions, give $r_{du} \approx 2 \times 10^{-8}$ duplications/gene/year.[Lynch, 2000] Substituting these values into (7.5) and then into (7.3) gives $M(n) \sim 10^{13}$-$10^{14}$ yrs, which is three to four orders of magnitude greater than the current estimate for the age of the Universe [Krauss, 2003]. Thus, the mean family formation times given by the linear BDIM would become realistic only if the recent analyses underestimated the gene duplication rate by a factor of $\sim 10^4$, which does not seem plausible. Accordingly, the linear BDIM cannot provide a realistic description of genome evolution, at least when only the mean time of family formation is considered. It can be shown that the variance of the family formation time is extremely large (the relative mean-square deviation is about 100), large deviations from the mean time, up to 2 orders of magnitude, are not improbable. In the next section, we consider non-linear, higher order models that have the potential to yield shorter mean times of family formation and in section 9 we turn to an alternative approach and describe simulations, which exploit the large number of families in evolving genomes and the substantial variance of the times of their formation.

## 8. Non-linear modifications of the BDIM

Theorem 1 asserts that a large class of models, namely the second order balanced BDIMs, provide any given power asymptotic of the stationary frequencies of family sizes. The linear BDIM is the simplest model that has the desirable asymptotics of stationary frequencies and fits well the real data. However, the more detailed analysis of the random process corresponding to the linear BDIM described above in sections 5, 6 and 7 showed that the characteristics of the stochastic behavior of the linear BDIM seem to be inconsistent with empirical data. The main problem is that the stochastic evolution of the linear BDIM is "too slow" and does not allow the formation of the large families that are actually observed in genomes.

Thus, our goal in this section is to modify the linear BDIM in such a way that:

i. the stationary distribution of the family sizes stays the same as for the linear BDIM;

ii. new models account for much more rapid evolution of family sizes for realistic values of duplication rates;

iii. the ratio of the mean times of family formation and extinction is substantially greater than it is under the linear model.

To provide for fast evolution of family sizes, the mean sojourn times $t_i$ in each state $i$, $t_i=1/(\lambda_i +\delta_i)$, should be substantially shorter then those in the linear model. The key mathematical point for the required modification is given by the following:

**Proposition 1**. Let $g(i)$, $i=0,…N$, be a positive function, $g(0)=1$. Consider the new transformed model (2.1) under simultaneous transformation of duplication and deletion rates given by the relations:

$$\lambda^*_i = \lambda_i\, g(i), \quad \delta^*_i = \delta_i\, g(i-1) \tag{8.1}$$

Then the stationary distribution for the BDIM with transformed birth and death rates, $\lambda^*_i$ and $\delta^*_i$, is the same as for the original model.

Note that the mean sojourn time of the modified model is $t^*_i=1/(\lambda_i\, g(i)+\delta_i\, g(i-1))$. Thus, $t^*_i$ can be arbitrarily decreased by choosing the appropriate function $g(i)$. It should be emphasized that the values of parameters $a$ and $b$ that have been previously determined for the linear BDIM to fit empirical data for different species (and Table 1) can be employed for the modified models because the transformation (8.1) does not change the stationary distribution. Theorem 1 indicates that models with rates (8.1) show the same asymptotic behavior as any polynomial model, although the initial behavior may differ.

We show that non-linear BDIM modifications with the function $g(i)=(i+1)^d$ satisfy the requirements (i) and (ii) and partially solve the problem (iii). In this section, we briefly describe the polynomial BDIMs of degrees $k=2$ and $k=3$ ($d=1$ and $d=2$, respectively). Other possible non-linear refinements will be discussed elsewhere.

Informally, polynomial BDIMs can be introduced as follows. Under the linear BDIM, the dependence of the birth and death rates on family size is very weak. This model actually does not include any form of interactions between family members, the growth rate is almost proportional to the family size and there is no significant feedback between the family size and growth rate. In contrast, the quadratic model includes dependence of birth and death rates of individual domains on pairwise interactions, whereas higher order models imply more complex interactions. In general, if interactions of the order $d$ are postulated, then the second order balanced BDIM has $\lambda_i \sim P_k(i)$ and $\delta_i \sim Q_k(i)$, where $P_k(i)$ and $Q_k(i)$ are polynomials on $i$ of the same degree $k = 1+d$ and the same higher coefficients. We show here that non-linear polynomial BDIM predict evolution rates that are dramatically greater then those for the linear BDIM and could be compatible with empirical estimates of duplication rates.

Let us consider the second order balanced *quadratic* BDIM with birth and death rates

$$\lambda_i = \lambda(i+a)(i+1),\ \delta_i = \lambda(i+b)i, \qquad (8.2)$$

i.e., the quadratic BDIM is a transformation of the linear BDIM with $g(i)=i+1$.

Similarly, the *cubic* BDIM has the birth and death rate as:

$$\lambda_i = \lambda(i+a)(i+1)^2,\ \delta_i = \lambda(i+b)i^2 \qquad (8.3)$$

and is a second order balanced BDIM resulting from the transformation of the linear BDIM by using the function $g(i)=(i+1)^2$. The equilibrium distributions for the quadratic and cubic BDIMs are exactly the same as that for the corresponding linear BDIM, but the stochastic properties of the higher order models are dramatically different from those of the linear model. The quadratic and particularly the cubic BDIMs display much more rapid evolution of genome size than the linear model with the same value of the parameter $\lambda$ (Tables 3 and 4 and Figs. 4, 6).

This brings the time required for the formation of families of the observed size closer to realistic spans. Specifically, with the empirical estimates of the duplication rates used above for the linear BDIM, the quadratic model gives the mean family formation times ~ $10^{11}$ yrs.

The stochastic behavior of the system and its characteristics also can be investigated in the broader framework of rational BDIMs. One would expect that increasing the degree (the "order of interaction") of the model should result in faster family evolution; this is, indeed, the case under a fixed value of the parameter $\lambda$ (Fig. 7). This plot clearly illustrates the dramatic acceleration of evolution with the increase of model degree.

However, there is a mathematical restriction that does not allow progressive shortening of the family formation time under this approach. Thus, although the cubic model results in a much shorter time measured in $1/\lambda$ than the quadratic model (compare Figs. 4 and 6), it also implies a much greater ratio $r_{du}/\lambda$, thus yielding similar time estimates in years under the given average rate of duplication $r_{du}$ (compare Tables 3 and 4).

Given this problem, we identified the model degrees yielding the minimum of the mean time of formation of the largest family for the rational BDIM with

$$\lambda_i = \lambda(i+a)(i+1)^d, \quad \delta_i = \lambda(i+b)i^d. \tag{8.4}$$

According to (7.2), the mean time of formation of a family of size $n$ from an essential singleton (for which extinction is not allowed) for this BDIM is

$$M(n, d) = 1/\lambda \, M_n(d) \tag{8.5}$$

where $M_n(d) = \sum_{m=1}^{n-1} \left( \frac{\Gamma(b+m+1)}{(m+1)^d \Gamma(a+m+1)} \sum_{k=1}^{m} \frac{\Gamma(a+k)}{\Gamma(b+k+1)} \right).$

Using the formal definition (7.4), we get $c_{du} = r_{du}/\lambda$, which connects the model parameter $\lambda$ with the empirically estimated duplication rate $r_{du}$:

$$c_{du}(d) = \left( \sum_{i=1}^{N-1} p_i \lambda_i / i \right)/\lambda = p_1 \left( \sum_{i=1}^{N-1} \frac{\Gamma(2+b)}{i\Gamma(1+a)} \frac{\Gamma(i+a+1)}{\Gamma(i+1+b)} (i+1)^d \right) =$$

$$\sum_{i=1}^{N-1} \frac{\Gamma(i+a+1)}{i\Gamma(i+1+b)} (i+1)^d / \sum_{i=1}^{N} \frac{\Gamma(a+i)}{\Gamma(b+i+1)} . \tag{8.6}$$

Let us denote

$$R_n(d) = c_{du}(d) M(n,d) = \tag{8.7}$$

$$[\sum_{i=1}^{N-1} \frac{\Gamma(i+a+1)}{i\Gamma(i+1+b)} (i+1)^d / \sum_{i=1}^{N} \frac{\Gamma(a+i)}{\Gamma(b+i+1)}] * [\sum_{j=1}^{n-1} (j+1)^{-d} \frac{\Gamma(b+k+1)}{\Gamma(a+k+1)} \sum_{i=1}^{k} \frac{\Gamma(a+i)}{\Gamma(b+i+1)}]$$

Then, excluding the model parameter $\lambda$ from (8.5) using (8.6), we get the following expression for the mean time of family formation, which is expressed through the mean duplication rate $r_{du}$, which was empirically estimated by Lynch and Conery [Lynch, 2000]:

$$M(n,d) = 1/r_{du} R_n(d). \tag{8.8}$$

Now we can see a notable effect: $R_n(d)$ with a set of species-specific parameters $a, b, N$ and a fixed $n$ has a minimum over $d$. Figure 8 and Table 5 show the dependence of $R_N(d)$, the time of formation of the largest family for the given species, on $d$. Due to this effect, increase of the model degree cannot yield family formation times $<10^{11}$ years (i.e., the minimal time is not dramatically smaller than that given by the quadratic model; Fig. 8), still not a realistic estimate.

## 9. Simulation of family evolution under BDIM of different degrees
In the previous section, we analytically determined the mean time of family formation for BDIM of different degrees and discovered that even the minimal mean time was substantially greater than the time allotted for genome evolution. However, for assessing the feasibility of the formation of the largest families during the evolution of real genomes, the more relevant value is not the mean but the minimum time of family formation over the entire ensemble of genes. Given the large variance of the family formation time estimates, this minimum value is likely to be much less than the mean.

Analytic determination of this value is hard so we resorted to Monte Carlo simulation analysis.

Evolution of an ensemble of families was simulated for the linear (3.2) and quadratic (8.2) BDIM. The model was initialized with 3000 families of size 1. At each discrete time step, for each family, a birth or death of a domain belonging to the family was simulated by (respectively) increasing or decreasing the family size counter; additionally, a new family of size 1 was created with the probability proportional to the innovation rate $\nu$ (although innovation was not explicitly used in the stochastic version of BDIM described in this work, it was incorporated into the simulations; the resulting process is analogous to the classic model of Karlin and McGregor [Karlin, 1967]). The probabilities of a birth or death for a given family of size $k$ were, respectively, proportional to $\lambda_k$ and $\delta_k$. The model parameters were estimated from the distribution of family sizes in human genome: $a$=5.16, $b$=6.43 (Table 1), $\nu/\lambda=\nu/\delta=2136.2$ [Karev, 2002]. The value of $\lambda=\delta$ was adjusted in such a way that the expected number of events of any type in a family during each model step was <0.1. For both linear and quadratic BDIM, the $c_{du}$ ratio was calculated according to (8.4); the time scale was adjusted in such a manner that $r_{du}=2\times10^{-8}$ duplications/gene/year [Lynch, 2000].

A series of simulations with each type of BDIM was run until the largest family reached the (arbitrary) size of 1024 members; the time (in years) when this happened was recorded (Fig. 9). For the linear BDIM, the median time required to produce the first family of size 1024 was 49.5 Ga (billion years) and the mean (±standard deviation) was 52.6±21.1 Ga. The quadratic BDIM reached this level much faster, with the median time of 2.52 Ga and the mean of 2.64±0.78 Ga. Perhaps not unexpectedly, these values are orders of magnitude smaller than the mean values estimated above. Thus, in these simulations, a realistic estimate of the time required for the formation of large families was reached for the quadratic but not for the linear BDIM. However, this should not be necessarily construed as a refutation of the latter given the large margin of error of the $r_{du}$ estimates.

## 10. Conclusions and perspective

In the previous work, we showed that the linear BDIM was the simplest in a broad class of birth, death and innovation models that gave a good fit to the empirically observed stationary distribution of domain family size for a variety of genomes. However, when explored in the stochastic regime, this model turned out to be inadequate, i.e., unable to account for sufficiently fast evolution to reach the observed family sizes given the time available for genome evolution and the best current estimates of duplication rates. Thus, we examined higher order degree BDIMs, which were obtained by a simple transformation of the linear model and generated the same stationary family size distribution. Models with degree between 2 and 3 yielded much more rapid evolution than the linear BDIM, which brings the characteristic times of family formation closer to realistic values, although the mean times of large family formation were still approximately two orders of magnitude greater than expected ($\sim 10^{11}$ yrs compared to the expected $\sim 10^9$ yrs). However, using Monte Carlo simulations, we showed that minimal time required for the formation of families of the expected size was much shorter than the mean time and, for the quadratic BDIM, was compatible with the actual time of genome evolution. Thus, the higher order BDIM are capable of producing realistic estimates of the family formation time. However, it would be premature to determine the order of BDIM that is optimal for the description of genome evolution because of the large margin of error on the empirical estimates of the duplication rates [Lynch, 2000].

Unlike the linear BDIM, higher order models imply interaction between family members, e.g., pairwise interactions in the case of the quadratic BDIM and third order interactions for the cubic BDIM. The interpretation of these interactions remains open. It does not seem likely that they should be rationalized as actual physical, functional or regulatory interactions. More realistically, these interactions could be thought of as a way to introduce into the model the positive selection pressure that drives the proliferation of paralogous families and accelerates it to such an extent that formation of the largest families observed in sequenced genomes becomes feasible. The adaptive significance of lineage-specific expansion of paralogous gene families has been considered in detail in comparative-genomic studies [Jordan, 2001; Lespinet, 2002; Remm, 2001]. Selection is the key element that is missing from the previous attempts on modeling genome evolution as a birth and death process and higher order BDIMs might be a means to bring it in "through the back door".

**Acknowledgements**

We thank Kasper Erikssen, Sergei Maslov, Dmitri Chklovsky, and Fyodor Kondrashov for useful discussions.

**Figures**

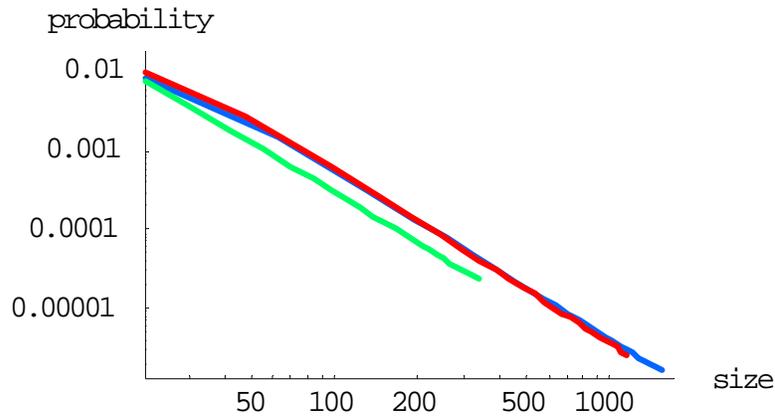

Fig. 1. Probabilities of formation of families starting from a singleton $P(1,n)$ versus the family size ($n$) for the linear BDIM. The plot is in double logarithmic scale. Species: *Drosophila melanogaster* (green), *Homo sapiens* (red), *Arabidopsis thaliana* (blue).

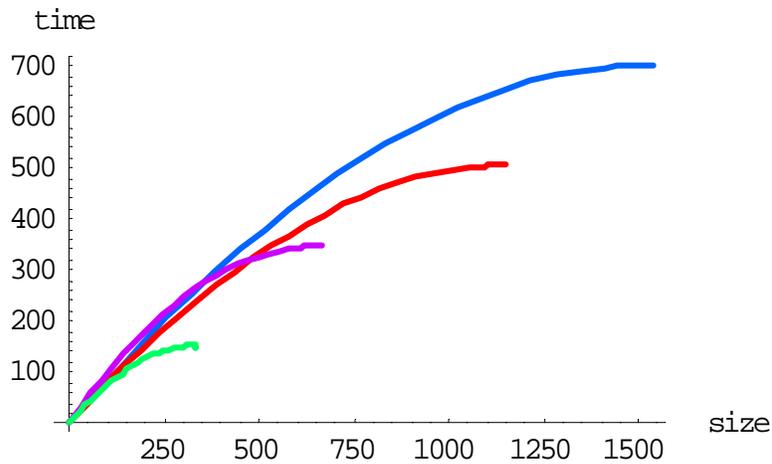

Fig. 2. The mean time of extinction ($E_n$) versus the family size ($n$) for the linear BDIM. Time is in $1/\lambda$ units. Species: *D. melanogaster* (green), *H. sapiens* (red), *A. thaliana* (blue), *C. elegans* (purple).

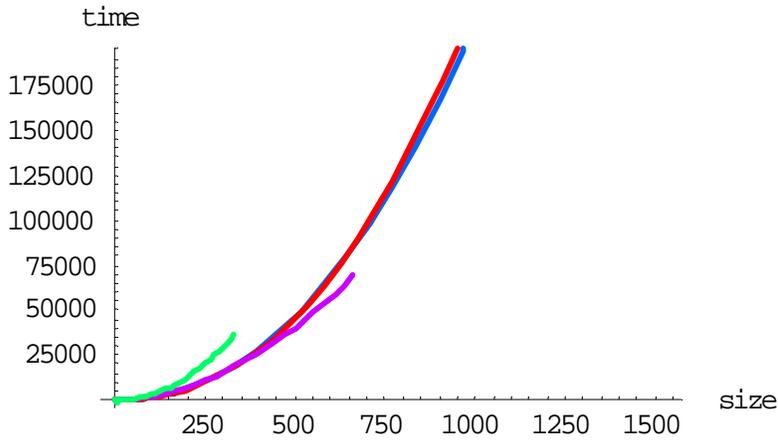

Fig. 3.a. Mean time of formation $M_n$ (in $1/\lambda$ units) versus family size ($n$) for the linear BDIM. Species: *D. melanogaster* (green), *H. sapiens* (red), *A. thaliana* (blue), *C. elegans* (purple).

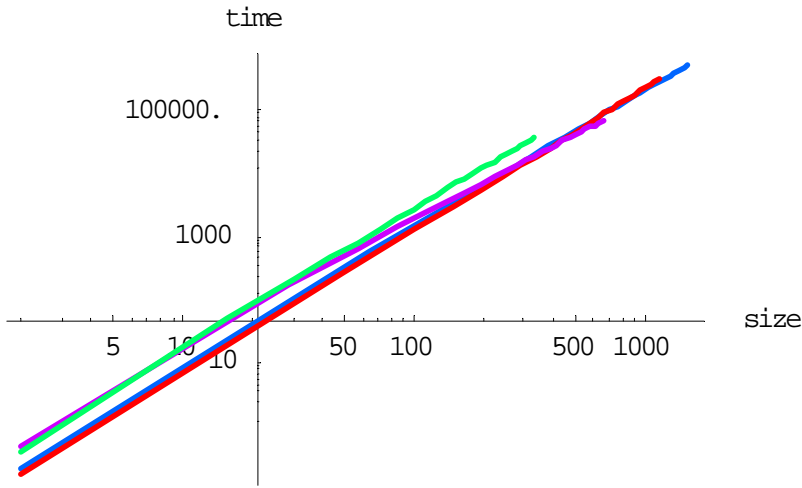

Fig. 3b. $M_n$ versus $n$ in double logarithmic scale for the linear BDIM.

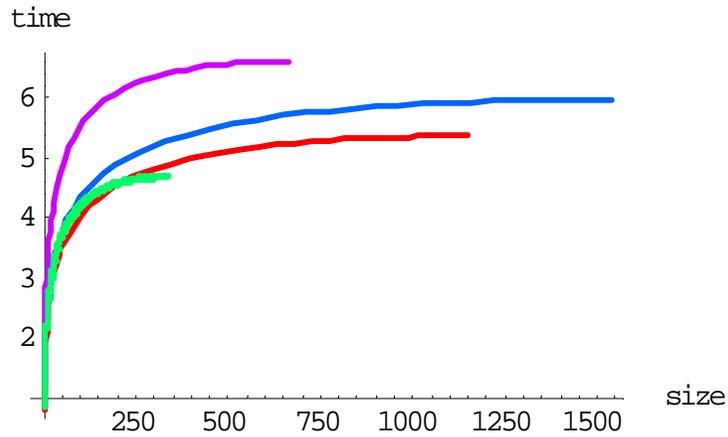

Fig. 4. Mean times of extinction $E_n$ (in $1/\lambda$ units) of family of size $n$ for the quadratic BDIM. Species: *D. melanogaster* (green), *H. sapiens* (red), *A. thaliana* (blue), *C. elegans* (purple).

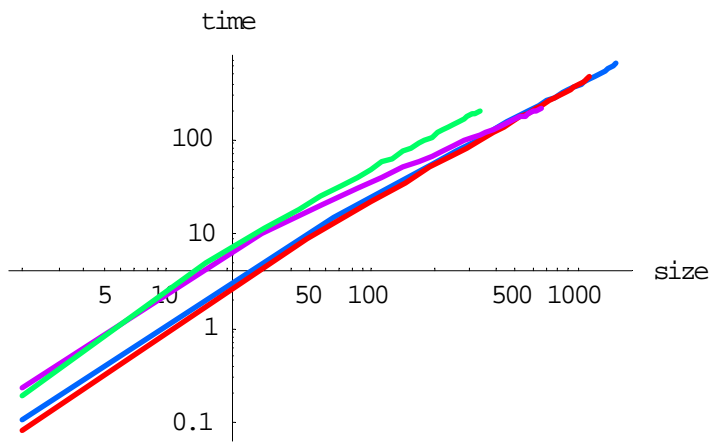

Fig. 5. Mean times of formation $M_n$ (in $1/\lambda$ units) of a family of size $n$ for the quadratic BDIM. The plot is in double logarithmic scale. Species: *D. melanogaster* (green), *H. sapiens* (red), *A. thaliana* (blue), *C. elegans* (purple).

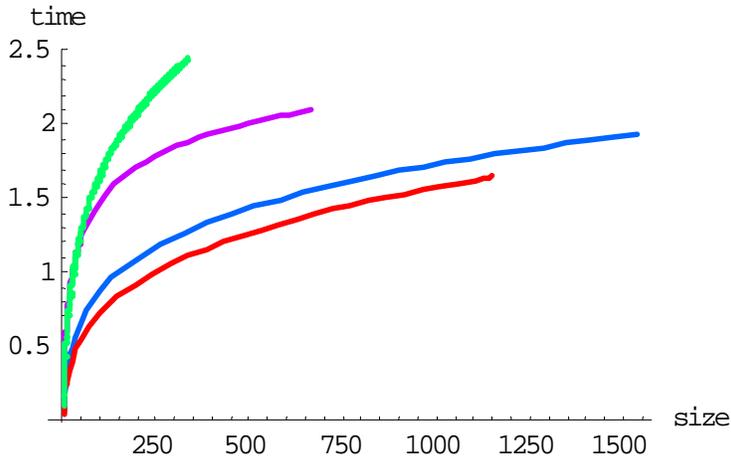

Fig. 6. Mean times of formation $M_n$ (in $1/\lambda$ units) of a family of size $n$ for the cubic BDIM. Species: *D. melanogaster* (green), *H. sapiens* (red), *A. thaliana* (blue), *C. elegans* (purple).

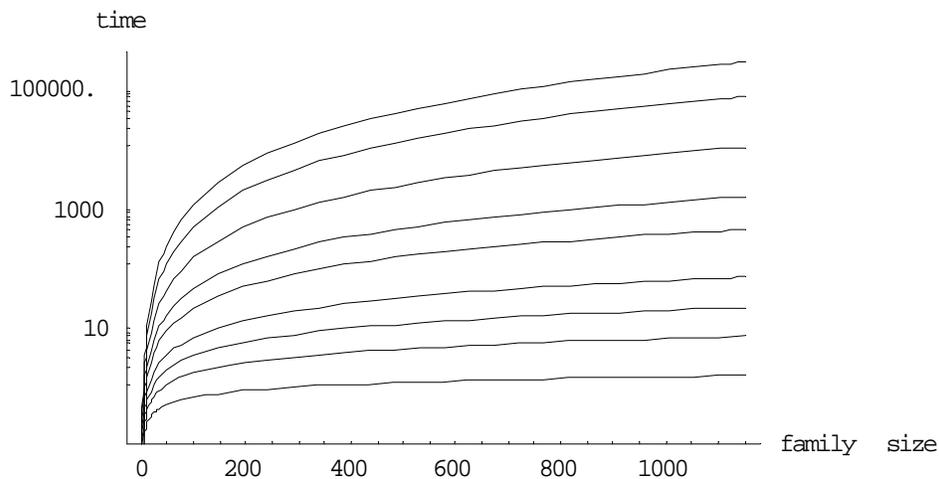

Fig. 7. The mean time of family formation for rational BDIMs. The plot shows the dependence of $M_n(d)$ on family size $n<N$ for non-linear BDIMs of different degrees $k=1+d$: $d=0$ (linear BDIM), 0.2, 0.5, 0.8, 1., 1.3, 1.5, 1.7, 2 (top to bottom). The values of parameters $a=5.16$, $b=6.43$, $N=1151$ are from the previous analysis of the *H. sapiens* genome ([Karev, 2002] and Table 1).

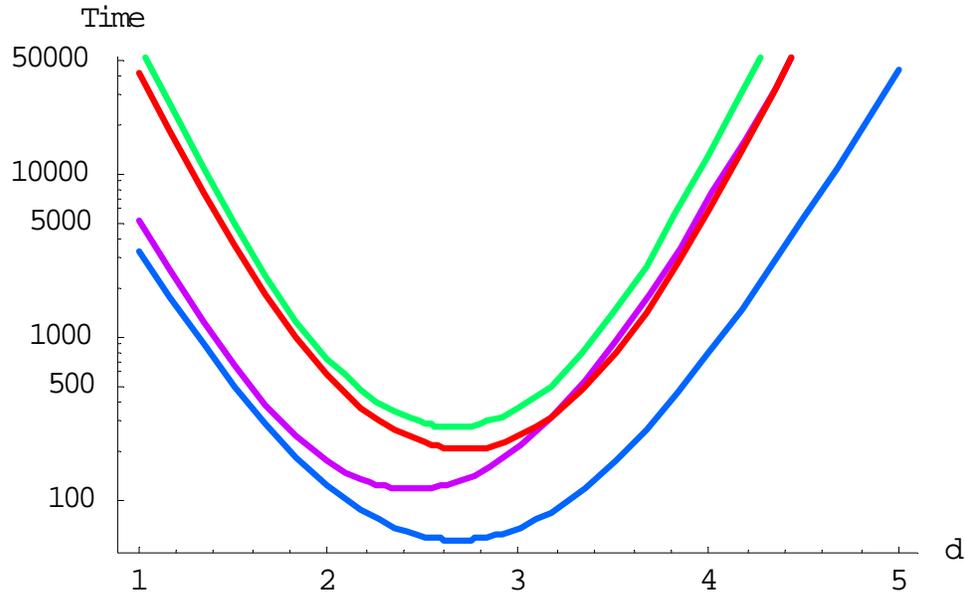

Fig. 8. Dependence of the time required for the formation of the largest family ($R_N(d)$) on the model degree $d$. Species: *D. melanogaster* (green), *H. sapiens* (red), *A. thaliana* (blue), *C. elegans* (purple).

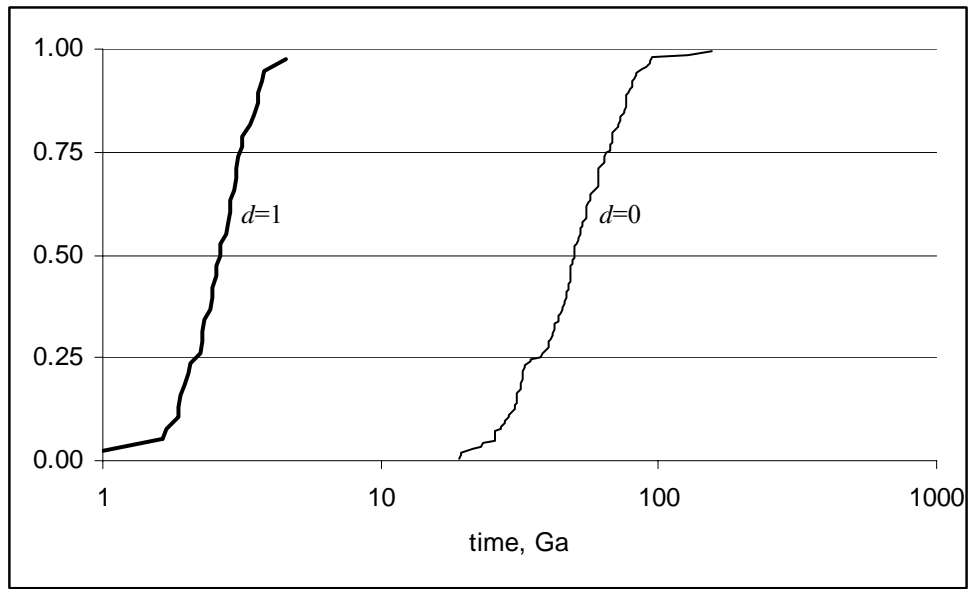

Fig. 9. Cumulative distribution functions of the time required for the formation of a family of size 1024 for the linear and quadratic BDIM.
The values of parameters $a=5.16$, $b=6.43$, $N=1151$ are from the previous analysis of the *H. sapiens* genome ([Karev, 2002] and Table 1).

**Table 1**

**Domain families in sequenced genomes and parameters of the best-fit linear BDIM**

| | No. of Protein-coding genes | No. of detected domain families | No. of detected domains | No. of proteins with RPS-BLAST Hits | Maximum size of a family | $a$ | $b$ | $\gamma$ |
|---|---|---|---|---|---|---|---|---|
| Sce[b] | 6340 | 1080 | 4575 | 3331 | 130 | 1.55 | 3.27 | 2.72 |
| Dme | 13605 | 1405 | 11734 | 7262 | 335 | 1.62 | 2.79 | 2.17 |
| Cel | 20524 | 1418 | 17054 | 11090 | 662 | 1.13 | 2.03 | 1.89 |
| Ath | 25854 | 1405 | 21238 | 15006 | 1535 | 3.80 | 4.98 | 2.18 |
| Hsa | 39883 | 1681 | 27844 | 16755 | 1151 | 5.16 | 6.43 | 2.27 |
| Tma | 1846 | 772 | 1683 | 1268 | 97 | 0.14 | 2.22 | 3.08 |
| Mth | 1869 | 693 | 1480 | 1150 | 43 | 0.12 | 2.00 | 2.88 |
| Sso | 2977 | 695 | 1950 | 1614 | 81 | 0.36 | 2.04 | 2.68 |
| Bsu | 4100 | 1002 | 3413 | 2502 | 124 | 0.48 | 2.01 | 2.53 |
| Eco | 4289 | 1078 | 3624 | 2765 | 140 | 0.84 | 2.54 | 2.70 |

**Table 2. Family evolution under the linear BDIM ($d=0$)**

|     | $N$ | $P_N(0)*10^4$ | $E_N(0)$ | $M_N(0)$ | $M_N(0)/E_N(0)$ | $c_{du}=r_{du}/\lambda$ | $M(N,0)$ |
| --- | --- | --- | --- | --- | --- | --- | --- |
| Sce | 130 | 0.284 | 47.46 | 20381.6 | 429.5 | 1.903 | 1939.3 |
| Dme | 335 | 0.227 | 153.74 | 37409.9 | 243.3 | 1.784 | 3337.0 |
| Cel | 662 | 0.160 | 347.76 | 68709.6 | 197.6 | 1.523 | 5232.2 |
| Ath | 1535 | 0.016 | 702.65 | 529639. | 753.8 | 2.382 | 63080.0 |
| Hsa | 1151 | 0.026 | 505.26 | 300665. | 595.1 | 2.721 | 40905.5 |
| Tma | 97 | 0.060 | 31.47 | 80677.3 | 2563.6 | 1.109 | 4473.6 |
| Mth | 43 | 1.125 | 14.91 | 4707.04 | 315.9 | 1.091 | 256.8 |
| Sso | 81 | 0.461 | 30.14 | 12853.5 | 426.5 | 1.253 | 805.3 |
| Bsu | 124 | 0.284 | 48.89 | 22921.0 | 468.8 | 1.320 | 1512.8 |
| Eco | 140 | 0.155 | 51.67 | 37959.8 | 734.7 | 1.544 | 2930.5 |

For the largest family of size $N$ in each genome, the table shows the probability of formation $P_N(d)$, mean times of formation $M_N(d)$ and extinction $E_N(d)$ ( in $1/\lambda$ units); mean times of formation $M(N,d)$ [in Ga ($10^9$ yrs), under $r_{du}=2\times 10^{-8}$].

**Table 3. Family evolution under the quadratic BDIM ($d=1$).**

|     | $N$ | $P_N(1)$ $*10^2$ | $E_N(1)$ | $M_N(1)$ | $M_N(1)/E_N(1)$ | $c_{du}=r_{du}/\lambda$ | $M(N,1)$ |
|-----|-----|------------------|----------|----------|-----------------|-------------------------|----------|
| Sce | 130  | 0.230 | 2.82 | 249.80  | 88.58  | 7.56  | 94.4  |
| Dme | 335  | 0.404 | 4.72 | 206.26  | 43.71  | 11.67 | 120.4 |
| Cel | 662  | 0.498 | 6.61 | 215.36  | 32.58  | 15.81 | 170.2 |
| Ath | 1535 | 0.131 | 5.98 | 638.27  | 106.73 | 22.50 | 718.1 |
| Has | 1151 | 0.166 | 5.37 | 468.84  | 87.31  | 24.48 | 573.9 |
| Tma | 97   | 0.039 | 2.25 | 1231.33 | 547.26 | 3.27  | 201.3 |
| Mth | 43   | 0.315 | 2.03 | 166.47  | 77.09  | 3.33  | 27.7  |
| Sso | 81   | 0.233 | 2.61 | 252.47  | 97.11  | 4.33  | 54.7  |
| Bsu | 124  | 0.212 | 3.10 | 304.97  | 98.38  | 5.09  | 77.6  |
| Eco | 140  | 0.135 | 2.90 | 431.85  | 148.91 | 5.74  | 123.9 |

For the largest family of size $N$ in each genome, the table shows the probability of formation $P_N(d)$, mean times of formation $M_N(d)$ and extinction $E_N(d)$ ( in $1/\lambda$ units); mean times of formation $M(N,d)$ (in Ga, under $r_{du}=2\times10^{-8}$).

**Table 4. Family evolution under the cubic BDIM ($d=2$).**

|     | $N$  | $P_N(2)$ | $E_N(2)$ | $M_N(2)$ | $M_N(2)/E_N(2)$ | $c_{du}=r_{du}/\lambda$ | $M(N,2)$ |
|-----|------|----------|----------|----------|-----------------|--------------------------|----------|
| Sce | 130  | 0.105    | 0.944    | 4.60     | 4.84            | 92.46                    | 21.3     |
| Dme | 335  | 0.222    | 1.390    | 2.45     | 1.76            | 549.65                   | 67.3     |
| Cel | 662  | 0.283    | 1.804    | 2.10     | 1.17            | 2020.37                  | 212.1    |
| Ath | 1535 | 0.255    | 1.390    | 1.93     | 1.39            | 3754.83                  | 362.3    |
| Hsa | 1151 | 0.254    | 1.291    | 1.65     | 1.27            | 2938.07                  | 242.4    |
| Tma | 97   | 0.019    | 0.781    | 24.48    | 31.4            | 18.84                    | 23.1     |
| Mth | 43   | 0.061    | 0.848    | 7.85     | 9.24            | 18.26                    | 7.2      |
| Sso | 81   | 0.073    | 0.960    | 7.21     | 7.51            | 36.71                    | 13.2     |
| Bsu | 124  | 0.088    | 1.059    | 6.40     | 6.04            | 63.38                    | 20.3     |
| Eco | 140  | 0.071    | 0.957    | 7.34     | 7.67            | 65.06                    | 23.9     |

For the largest family of size $N$ in each genome, the table shows the probability of formation $P_N(d)$, mean times of formation $M_N(d)$ and extinction $E_N(d)$ ( in $1/\lambda$ units); mean times of formation $M(N,d)$ (in Ga, under $r_{du}=2\times10^{-8}$).

**Table 5. Rational BDIM yielding the shortest time of family formation.**

|     | N    | d    | $R_N(d)$ | $M(N, d)$ |
|-----|------|------|----------|-----------|
| Sce | 130  | 2.13 | 416.0    | 20.8      |
| Dme | 335  | 1.67 | 1131.0   | 56.55     |
| Cel | 662  | 1.44 | 2321.4   | 116.1     |
| Ath | 1535 | 1.65 | 5553.8   | 277.7     |
| Hsa | 1151 | 1.71 | 4079.5   | 204.      |
| Tma | 97   | 2.56 | 317.8    | 15.9      |
| Mth | 43   | 2.40 | 125.2    | 6.3       |
| Sso | 81   | 2.19 | 254.2    | 12.7      |
| Bsu | 124  | 2.05 | 404.4    | 20.       |
| Eco | 140  | 2.16 | 460.4    | 23.       |

For each genome, the value of $d$, which results in the minimum of the mean time of formation of the largest family, $M(d, N) = R_N(d)/r_{du}$ (in Ga), is indicated.